\pdfoutput=1
\documentclass[12pt]{article}
\usepackage[a4paper,
            mag=1000, includefoot,
            left=3cm, right=1.5cm, top=2cm, bottom=2cm, headsep=1cm, footskip=1cm]{geometry}
\usepackage{mathptmx}

\usepackage{helvet}
\usepackage{courier}
\usepackage{type1cm}

\usepackage{makeidx}         % allows index generation
\usepackage{graphicx}        % standard LaTeX graphics tool
\usepackage[bottom]{footmisc}% places footnotes at page bottom

\usepackage[english]{babel}

\usepackage{latexsym,amssymb,amsbsy}
\usepackage{amsthm}
\usepackage{amsmath}
\usepackage{siunitx}
\usepackage{morefloats}
\usepackage[section]{placeins}
\usepackage{lscape}
\usepackage{longtable}
\usepackage{caption}
\usepackage{subcaption}

\usepackage{graphics}
\usepackage{tikz}
\usepackage{ifpdf}
\ifpdf\usepackage{epstopdf}\fi
\usepackage{url}
\usepackage{bm}
\usepackage{mathtools}
\usepackage{xcolor}
\usepackage{colortbl}
\usepackage{hyperref}

\def\spN{\mathbb{N}}
\def\spR{\mathbb{R}}
\def\spZ{\mathbb{Z}}





\usepackage{eucal}

\newcommand{\rmT}{\mathrm{T}}

\newcommand{\tN}{\textsf{N}}

\newcommand{\tP}{\textsf{P}}

\newcommand{\tS}{\textsf{S}}
\newcommand{\tT}{\textsf{T}}

\newcommand{\tX}{\textsf{X}}
\newcommand{\tY}{\textsf{Y}}

\newcommand{\bfX}{\mathbf{X}}

\newcommand{\bt}{\begin{theorem}}
\newcommand{\et}{\end{theorem}}
\newcommand{\bl}{\begin{lemma}}
\newcommand{\el}{\end{lemma}}
\newcommand{\bp}{\begin{proposition}}
\newcommand{\ep}{\end{proposition}}
\newcommand{\bc}{\begin{corollary}}
\newcommand{\ec}{\end{corollary}}

\newcommand{\bd}{\begin{definition}\rm}
\newcommand{\ed}{\end{definition}}
\newcommand{\bex}{\begin{example}\rm}
\newcommand{\eex}{\end{example}}
\newcommand{\br}{\begin{remark}\rm}
\newcommand{\er}{\end{remark}}

\newcommand{\btbh}{\begin{table}[!htbp]}
\newcommand{\etb}{\end{table}}
\newcommand{\bfgh}{\begin{figure}[!htbp]}
\newcommand{\efg}{\end{figure}}

\newcommand{\bea}{\begin{eqnarray*}}
\newcommand{\eea}{\end{eqnarray*}}
\newcommand{\be}{\begin{eqnarray}}
\newcommand{\ee}{\end{eqnarray}}

\newcommand{\suml}{\sum\limits}

\newcommand{\vphi}{\varphi}

\newcommand{\lm}{\lambda}

\newcommand\Var{\mathrm{D}}

\DeclareMathAlphabet\mathbfcal{OMS}{cmsy}{b}{n}

\makeatletter
\def\adots{\mathinner{\mkern2mu\raise\p@\hbox{.}
\mkern2mu\raise4\p@\hbox{.}\mkern1mu
\raise7\p@\vbox{\kern7\p@\hbox{.}}\mkern1mu}}
\newcommand{\l@abcd}[2]{\hbox to\textwidth{#1\dotfill #2}}
\makeatother

\def\unit{\mathfrak{i}}

\usepackage[shortlabels]{enumitem}
\newlist{arguments}{description}{1}
\setlist[arguments]{style=sameline}
\newlist{inlinelist}{enumerate*}{1}
\setlist*[inlinelist,1]{label=(\alph*), itemjoin={{, }}, itemjoin*={{, and }}}

\usepackage{algpseudocode}
\usepackage{capt-of}

\algnewcommand\algorithmicinput{\textit{Input:}}
\algnewcommand\Input{\item[\algorithmicinput]}
\algnewcommand\algorithmicoutput{\textit{Output:}}
\algnewcommand\Output{\item[\algorithmicoutput]}
\newcommand{\algrule}[1][1pt]{\par\vskip.5\baselineskip\hrule height #1\par\vskip.5\baselineskip}

\newcounter{ssaalg}[section]
\makeatletter
\renewcommand\thessaalg{\thesection.\@arabic\c@ssaalg}%
\def\ext@ssaalg{loa}
\def\fnum@ssaalg{\textsc{Algorithm~\thessaalg}}
\makeatother
\newenvironment{algorithm}[2]{%
    \begingroup
    \vskip 2 em
    \par\nobreak\vskip-.5\baselineskip\hrule height 1pt\par\nobreak\vskip.5\baselineskip\nopagebreak%
    \captionof{ssaalg}{#1\label{#2}}\nopagebreak%
    \begin{algorithmic}[1]%
    \par\nobreak\vskip-.5\baselineskip\hrule height 1pt\par\nobreak\vskip.5\baselineskip\nobreak%
    \addtolength{\itemsep}{2pt}}{%
    \nopagebreak\algrule%
    \end{algorithmic}%
    \vskip 1em%
    \endgroup}

\makeatletter
\newcommand\code{\bgroup\@makeother\_\@makeother\~\@makeother\$\@codex}
\def\@codex#1{{\small\ttfamily\hyphenchar\font=-1 #1}\egroup}
\makeatother

\usepackage{xspace}
\def\R{{\normalfont\ttfamily R}\xspace}

\makeatletter
\def\verbatim@font{\small\ttfamily}
\makeatother

\usepackage{latexsym, amsmath, amssymb, amsthm}
\usepackage{tikz}
\usepackage{microtype}

\DeclareMathOperator*{\argmax}{arg\,max}
\DeclareMathOperator*{\argmin}{arg\,min}

\graphicspath{{fig/}}

\newtheorem{definition}{Definition}
\newtheorem{corollary}{Corollary}
\newtheorem{theorem}{Theorem}
\newtheorem{remark}{Remark}
\newtheorem{proposition}{Proposition}
\newtheorem{lemma}{Lemma}
\theoremstyle{definition}
\newtheorem{example}{Example}

\usepackage{xifthen}% provides \isempty test

\newcommand{\optarg}[1][]{%
  \ifthenelse{\isempty{#1}}%
    {}% if #1 is empty
    {(((#1)))}% if #1 is not empty
}

\begin{document}

\author{Nina Golyandina\footnote{St.Petersburg State University, Universitetskaya nab. 7/9, St.Petersburg, Russia. n.golyandina@spbu.ru}, Polina Zhornikova\footnote{St.Petersburg State University, Universitetskaya nab. 7/9, St.Petersburg, Russia. polina.zhornikova@gmail.com}
%\footnote{The work is supported by the Russian Science Foundation (project No. 23-21-00222).}
}
\title{On automated identification in singular spectrum analysis for different types of objects}
%\subtitle{-- Monograph --}
\date{}
\maketitle

\begin{abstract} Approaches to automated grouping in singular spectrum analysis are considered. A new method for the identification of periodic components is proposed. The possibilities of extensions to multivariate time series and images are discussed.
\end{abstract}

\tableofcontents

\section*{Introduction}

In this paper, we consider the problem of automation of singular spectrum analysis (SSA) for time series decomposition into a trend, periodical components and noise.

Let us briefly describe the SSA algorithm (see \cite{Golyandina.Zhigljavsky2020book} for details), which is rapidly developed and is used for solving many real-life problems.
First, the time series $\tX$ is transformed to the so-called trajectory matrix $\bfX$ of size $L\times K$. Then the singular value decomposition $\bfX = \sum \sqrt{\lambda_i} U_i V_i^\rmT$ is constructed. This decomposition consists of elementary matrices of rank one. The most sophisticated part of the SSA algorithm is to identify and then gather the elementary matrices into groups corresponding to trend and periodic components. After identification, the elementary matrix components are summed and then transforms back to time series by diagonal averaging.
There are several papers devoted to automation of the grouping step \cite{Vautard.etal1992,Alexandrov.Golyandina2004en,Alexandrov.Golyandina2005,Alexandrov2006en,Alexandrov2009,AlvarezMeza.etal2013,Harmouche.etal2018,Kalantari.Hassani2019,Jain.etal2020,Bogalo.etal2021}.
Implementation in \R{} of several methods is described in \cite{Golyandina.etal2018}.

We consider the approach, which is based on the following property: the behaviour of elementary time series, eigenvectors $U_i$ and factor vectors $V_i$ repeats the behaviour of the time series component that produces them. Extraction of trend and quasi-periodic components composed of exponentially-modulated harmonics is studied.

In Section~\ref{sec:auto1D}, we review the methods for one-dimensional series and propose a new method for the identification of harmonics (Section~\ref{sec:tau1}). In Sections~\ref{sec:auto_mssa} and \ref{sec:auto2D} we consider possible extensions to the decomposition of multivariate time series and 2D digital images.

Appendix in Section~\ref{sec:append} contains a description of the necessary properties of the SSA decomposition for exponentially-modulated harmonics.

\section{Automated grouping for time series}
\label{sec:auto1D}

In this section, we will consider the problem of automatic identification of decomposition components in SSA. Let us start with the definition of the components of the time series that we want to identify.

Consider a time series $\tX= (x_1,\ldots,x_N)$, $x_i \in \spR$.

We call a low-frequency component \textit{trend}. For a one-dimensional real-valued time series, this is the component $\tT$, for which in the Fourier expansion of the series $\tT$
\begin{gather*}
t_n = C_0 + \sum_{k=1}^{\lfloor (N-1)/2 \rfloor}\sqrt{C_k^2 + S_k^2} \cos(2\pi n k /N + \phi_k) + C_{N/2} (-1)^n
\end{gather*}
 the largest values have the coefficients $\sqrt{C_k^2 + S_k^2}$ with a small value of $k$; the last summand $C_{N/2} (-1)^n$ is present only if $N$ is even. %For the other kinds of object $\tX$ a formal definition of the trend will be given later in the paper.

Various regular oscillations will be called \textit{oscillatory components}. Formally, we will consider a sum of exponentially-modulated (e-m) harmonic series (harmonics).
For a one-dimensional real series, the $n$th element of an e-m harmonic with frequency $\omega$ ($\omega \leq
0.5$) is given by the expression:
$
a \, e^{\alpha n} \cos(2\pi \omega n + \phi)
$, $0 \leq \phi < 2\pi$, $a \not = 0$.
 %For the other kinds of object $\tX$, the formal definition of the e-m harmonic will be given in the paper.

A sum of the trend and the oscillatory components will be called \textit{signal}.

We assume that the time series $\tX$ contains the following additive components: trend $\tT$, oscillatory component $\tP$ and random noise $\tN$. Thus, in general, the considered model of the object $\tX$ looks like this:
\begin{gather*}
\tX = \tT+\tP+\tN,
\end{gather*}
where the elements of $\tN$ are realizations of a random variable.

\subsection{Low-frequency method for trend identification}
\label{sec:1d_freq_method}

The considered automation method was introduced in \cite{Alexandrov.Golyandina2004en}. An approach to the automatic selection of parameters of the automation method is proposed in \cite{Alexandrov2006en}; we will not consider it here. The automation method allows the identification of the components related to the trend.
 We will call this method the \textit{low-frequency method for trend identification}.

  For a series $\tY$ of length $M$ and

Let us introduce the periodogram:
\begin{gather}
\label{eq:per}
 \Pi_{\tY}^M(k/M) = \frac{M}{2}
\begin{cases}
2C_0^2 &\quad \text{for } k = 0, \\
C_k^2 + S_k^2 & \quad \text{for }  0 \leq k \leq M/2, \\
2C_{M/2}^2 & \quad \text{for } k = M/2   \text{, if } M \text{ even},
\end{cases}
\end{gather}
where the coefficients $C_k$ and $S_k$ are taken from the Fourier
decomposition of $\tY=(y_1,\ldots,y_M)$:
\bea
\label{eq:FOUR2}
   y_n = C_0 +
  \suml_{k=1}^{\lfloor M/2 \rfloor} \Big(C_k\cos(2\pi n\,k/M) + S_k \sin(2\pi n\,k/M)\Big).
\eea

For a series $\tY$ of length $M$ and
for $0 \le \omega_1 \le \omega_2\le 0.5$, we define
\be
\label{eq:freq_share}
T(\tY; \omega_1, \omega_2)=\sum_{k: \omega_1\leq k/M <\omega_2}I_\tY^M(k/M),
\ee
where
\be
\label{eq:per_I}
I_\tY^M(k/M)=\Pi_\tY^M(k/M)/\|\tY\|^2,
\ee
$\Pi_\tY^M$ is defined in \eqref{eq:per}.
Since we have $\|\tY\|^2 = \sum_{k=1}^{[M/2]} \Pi_\tY^M(k/M)$,
the measure $T(\tY; \omega_1, \omega_2)$ can be
considered as a proportion of frequencies contained in
the frequency bin $[\omega_1, \omega_2)$.

One of the aims in performing grouping is the extraction of a series component with
frequency range mostly from the chosen frequency bin. Therefore, it is natural to calculate the value
of $T$ for elementary reconstructed components.
Moreover, SSA reconstruction can be considered as a linear filter.
It appears that the frequency response of the filter generated by the $i$th eigentriple
is almost the same as the periodogram of the corresponding singular vector, see
\cite[Proposition 3.13]{Golyandina.Zhigljavsky2020book}.
Therefore, it is reasonable to apply $T$ also to singular vectors to reconstruct the series components
with the given frequency ranges.

Since the trend of a series can be defined as its slowly varying series component, for extracting a trend,
a frequency bin in the form $[0,\omega)$ should be chosen.
Then we consider
\be
\label{eq:T_measure}
T(\tY; \omega)=\sum_{k: k/M <\omega}I_\tY^M(k/M).
\ee

The value of $\omega$ reflects the frequency range, which we associated with a trend. For example, if the series has monthly seasonality,
$\omega$ should be notably smaller than $1/12$.
Note that the grouping method does not answer the question of whether the extracted component is indeed a deterministic trend or simply a result of smoothing.

Values of $T$ for each elementary decomposition component can be used for performing the grouping.
To perform an automatic grouping, a threshold $T_0$, $0\le T_0 \le 1$, should be given.
For example, if the value $T(\tY_i; 0, \omega)$ is larger than $T_0$ for some small $\omega$, where
$\tY_i$ is the $i$th elementary series or $i$th left/right singular vector,
then the corresponding eigentriple can be automatically considered as a part of the trend.

The suggested method is presented in Algorithm~\ref{alg:freq1d}.

\begin{algorithm}{1D-SSA: Frequency identification of trend components, by the threshold}{alg:freq1d}
    \Input{Frequency range $[\omega_1,\omega_2)$, threshold $T_0$, group $I$,
    type of series: eigenvectors, factor vectors or reconstructed series.}
    \Output{A group of components $J\subset I$.}

    \State
    For each series $\tY_i$, $i\in I$, the measure $T(\tY_i; \omega_1, \omega_2)$ given in \eqref{eq:freq_share}
    is calculated.
     \State
    The resultant group $J$ consists of indices $i\in I$ such that $T(\tY_i; \omega_1, \omega_2)\ge T_0$.
\end{algorithm}

\subsection{Frequency method for identifying the oscillating component}
\label{sec:1d_pgram_method}

Here we define the oscillatory component as a sum of the e-m harmonics.
Let us consider the algorithm for the automatic identification of the e-m harmonics. %proposed in \cite{Alexandrov.Golyandina2005}.
 We will call this method \textit{frequency method of identification of the oscillatory component}.

The method was suggested in \cite{Vautard.etal1992} and further developed in \cite{Alexandrov.Golyandina2005} and \cite{Alexandrov2006en}; it is based on the study of periodograms of singular vectors corresponding to the series. The method consists of two parts, at the first stage a preliminary check is carried out, and the singular triples identified at this stage are further checked at the second stage.

As we know from Proposition \ref{th:ex_mod_1d}, an e-m harmonic \eqref{eq:ex_mod_garm_1d} can have rank 1 if frequency $\omega=0.5$, or rank 2 otherwise, i.e. the e-m harmonic can correspond to either one singular vector or two.

The considered method for automatically identifying the components corresponding to e-m harmonics checks
\begin{itemize}
\item each pair of singular vectors if they are similar to e-m harmonics with the same frequency,
\item each singular vector if it is similar to an e-m harmonic with period 2 ($\omega=0.5$).
\end{itemize}

\begin{remark}
Since under some non-restrictive conditions a one-dimensional real-valued harmonic series with $\omega < 0.5$ produces two equal (Proposition~\ref{th:1dssa_num}) or close (\cite{Golyandina.etal2001}) eigenvalues of the trajectory matrix and the SSA method algorithm sorts components of the SVD by eigenvalues,
it is sufficient to consider only consecutive pairs of singular vectors to identify harmonics of rank 2.
\end{remark}

According to Section \ref{sec:1d_ssa_theory}, a harmonic series of the form \eqref{eq:ex_mod_garm_1d} corresponds to two singular vectors of the form \eqref{eq:vectors_ex_mod_1d} with equal frequencies.

The following result is known; the normalized periodogram $I_\tY^M$ is given by \eqref{eq:per_I}.

\begin{proposition}  \cite[Proposition 3.1]{Alexandrov2006en} \label{th:aleks1}
Let $\alpha = 0$, $\omega < 0.5$ and $L\omega \in \spN$. Then for singular vectors $U_1$, $U_2$ of time series $\tS$ with elements given by \eqref{eq:ex_mod_garm_1d}
\begin{gather*}
	\max_{0 \leq k \leq L} I_{U_1}^L(k/L) = \max_{0 \leq k \leq L} I_{U_2}^L(k/L) = I_{U_1}^L(\omega) = I_{U_2}^L(\omega) = 1.
\end{gather*}
\end{proposition}

The first step of the method is based on the fact that if for a harmonic ($\alpha = 0$) with frequency $\omega$ and window length $L$ the relation $L \omega \in \spN$ is fulfilled, then the periodograms of its singular vectors are $I_{U_j}^L(k/L)=\chi_{\omega}(k/L)$ (where $\chi_{\omega}(\cdot)$ --- indicator of set $\{\omega\}$).
Given only approximate separability when $L\omega \not \in \spN$ or $\alpha \not = 0$,
at the first stage of the method, we choose from all consecutive pairs of singular vectors those for which arguments of periodogram maxima are greater than 0 and close, i.e. differ from each other by no more than $s_0/L$ (where $s_0 \in \spZ_+$ is a fixed parameter of the method):
\begin{gather} \label{eq:I_1_P}
J_1^{(\tP)} = \{ (i, i+1): \quad \theta_1, \theta_2 >0, \quad L |\theta_i - \theta_{i+1}| \leq s_0, \quad 1 \leq i \leq d -1 \},
\end{gather}
where $\theta_j = \argmax_{0 < k \leq L/2} \{I_{U_j}^L(k/L)\}$ is the argument of the maximum of the periodogram $I_{U_j}^{L}$ of the singular vector $U_j$.

Similarly, each singular vector is checked against a harmonic with a frequency of 0.5:
\begin{gather} \label{eq:I_2_P}
J_2^{(\tP)} = \{i: \quad L |\theta_{i} - 0.5 | \leq s_0, \quad 1 \leq i \leq d\}.
\end{gather}

The results of the first step are the sets $J_1^{(\tP)}$ and $J_2^{(\tP)}$ of the singular vectors' numbers. The set $J_1^{(\tP)}$ consists of pairs of numbers of singular vectors identified as corresponding to e-m harmonics with $0 < \omega < 0.5$.  The set $J_2^{(\tP)}$ contains numbers of singular vectors identified as corresponding to e-m harmonics with $\omega = 0.5$.

\begin{definition} \label{def:rho}
Let ${\cal{A}}= \{W_j\}$, $W_j$ be a finite set of real vector vectors, denote the power of the set as $\# {\cal{A}}$. Let us introduce a function $\rho_{\cal{A}}$ defined on the set ${\cal{A}}$:
\begin{gather*}
\rho_{{\cal{A}}}(k/L) = \frac{1}{\#{\cal{A}}} \sum_{W_j \in {\cal{A}}}{I^{L}_{W_j}{(k/L)}}
\end{gather*}
\end{definition}

In view of Proposition \ref{th:aleks1}, to check that the singular vectors with numbers $j, j + 1$ correspond to a harmonic, we must ensure that their periodogram maxima are reached at one point and that each of them has a value close to 1, since e-m harmonic with frequency $\omega$ such that $L \omega \not \in \spN$ also corresponds to a large value of the periodogram maximum (though smaller than 1), because its periodogram has one distinct peak.  This is what is checked in the second step of the method.

Before \cite{Alexandrov.Golyandina2005}, based on which we describe the method, the considered approach was proposed in \cite{Vautard.etal1992}, but the idea itself is not new and was proposed back in 1929 by Fisher in \cite{Fisher1929}.

Thus, the second step of the method
 is to choose a threshold $\rho_0$ for the measure
\begin{gather} \label{eq:rho_ij}
\rho_{i,j} := \max_{0 < k \leq L/2}{\left(\rho_{\{{U_i,U_j\}}(k/L) + \rho_{\{U_i,U_j\}}}((k+1)/L)\right)},
\end{gather}
where $U_i$, $U_j$ are a pair of singular vectors with indices from the set of indices $J_1^{(\tP)}$ selected in the first step.

To identify the components belonging to a harmonic of frequency 0.5, introduce the measure in the form
\begin{gather}\label{eq:rho_i}
\rho_{i} := \rho_{\{U_i\}}((\lfloor L/2 \rfloor)/L) + \rho_{\{U_i\}}((\lfloor L/2 \rfloor + 1)/L),
\end{gather}
where $i \in J_2^{(\tP)}$.

The final result of the frequency method implemented by Algorithm~\ref{alg:1d_pgram} is the indices
\begin{gather} \label{eq:pgram_I_p}
J^{(\tP)} = \{ (i,j) \in J_1^{(\tP)}: \rho_{i,j} \geq\rho_0 \} \cup \{ i \in J_2^{(\tP)}: \rho_{i} \geq\rho_0 \}.
\end{gather}

\begin{algorithm}{1D-SSA. Frequency identification of oscillatory components}{alg:1d_pgram}
\Input The data and parameters are as follows.
\begin{enumerate}
\item \textbf{Data}: left singular vectors $\{U_i\}_{i=1}^{d}$.
\item \textbf{Parameters}: parameter $s_0 \in \spZ_{+}$, threshold $\rho_0 \in [0,1]$.
\end{enumerate}
\Output A group of component indices $J^{(\tP)}$ related to the oscillatory component.
\State Based on  $\{U_i\}_{i=1}^{d}$, we obtain the index group $J_1^{(\tP)}$ using \eqref{eq:I_1_P} with $s_0$ and the index group $J_2^{(\tP)}$ using \eqref{eq:I_2_P} with $s_0$.
\State Obtain the index group $J^{(\tP)}$ using \eqref{eq:pgram_I_p}  with $\rho_0$ applied to $\{U_j\}_{j \in J_1^{(\tP)} \cup J_2^{(\tP)}}$.
\end{algorithm}

The author \cite{Alexandrov2006en} advises taking $s_0 = 1$. The disadvantage of the method is the fact that it is originally designed to identify unmodulated harmonics. When applied to simulated harmonics, the threshold \cite{Alexandrov.Golyandina2005} must be reduced, which can lead to false detection of harmonics in the unmodulated case ($\alpha = 0$).

\subsection{Method for identifying the oscillatory component by the regularity of angles}
\label{sec:tau1}
\subsubsection{Description and justification}

Let $P = (p_1,\ldots,p_L)^{\mathrm{T}}$ and $Q=(q_1,\ldots,q_L)^{\mathrm{T}}$ be two real vectors of length $L$. Introduce the measure
\begin{gather} \label{eq:tau1}
\tau(P, Q) := \hat{\Var}(\Theta) =\frac{1}{L-1}\sum_{k=1}^{L-1}{\left(\theta_k  - \bar{\theta}\right)^2},
\end{gather}
where $\Theta=(\theta_1,\ldots,\theta_L)^{\mathrm{T}}$,
$\bar{\theta} = \sum_{k=1}^{L-1}{\theta_k}/(L-1)$, $\theta_k$ is the angle between
$\left(p_k, q_{k}\right)^{\mathrm{T}}$ and $\left(p_{k+1}, q_{k+1}\right)^{\mathrm{T}}$, that is,
\begin{gather*}
\theta_k = \arccos{\left(\frac{p_{k} p_{k+1} + q_{k} q_{k+1}}{\sqrt{p_{k}^2+q_{k}^2}\sqrt{p_{k+1}^2+q_{k+1}^2}}\right)}.
\end{gather*}
Note that the values of $\theta_k$ belong to $[0,\pi]$.

\begin{proposition}  \label{th:th_tau_1d}
For singular vectors $U_1$ and $U_2$  of the time series $\tS$ with elements given by \eqref{eq:vectors_ex_mod_1d}, the following statements are valid.
\begin{enumerate}
\item \label{th:re_tau1}
If $\alpha=0$ and $L\omega$ is integer, then $\tau(U_1, U_2)=0$.
\item \label{th:re_tau2}
If $\alpha = 0$ and $L=[\beta N]$, where $0<\beta<1$, then $\lim_{L \rightarrow \infty}\tau(U_1, U_2) = 0$.
\item  \label{th:re_tau3}
If $\alpha = \alpha_N = C/N$, where $C$ is some constant, and $L=[\beta N]$, where $0<\beta<1$, then $\lim_{L \rightarrow \infty}\tau(U_1, U_2) = 0$.
\end{enumerate}
\end{proposition}

Let us describe the approach to identifying an oscillating component based on the properties of the measure $\tau$.
The equality $\tau(P, Q) =0$ means that the angles between the sequential points $(p_k, q_k)^{\mathrm{T}}$ are equal. If the norms of the vectors $P$ and $Q$ are monotonically changed in $k$, then the 2D diagrams of $P$ and $Q$ look like spirals.

\begin{figure}[h]
    \centering
   \begin{subfigure}[b]{0.3\textwidth}
       \includegraphics[width=\textwidth]{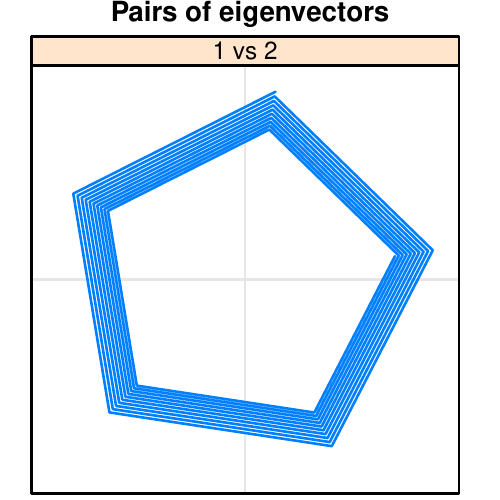}
        \caption{$\omega=1/5$, $\alpha = 0.005$, \\$C = 0.495$, $\tau =1.3e-05$.}

    \end{subfigure}
    \begin{subfigure}[b]{0.3\textwidth}
        \includegraphics[width=\textwidth]{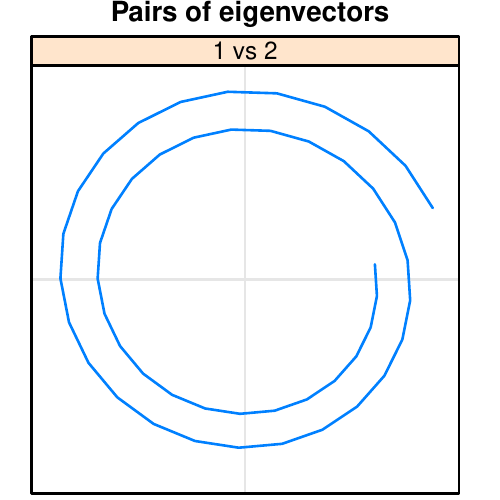}
        \caption{ $\omega=1/25$, $\alpha = 0.009$, $C = 0.891$, $\tau =$4.2e-05.}
    \end{subfigure}
    \begin{subfigure}[b]{0.3\textwidth}
        \includegraphics[width=\textwidth]{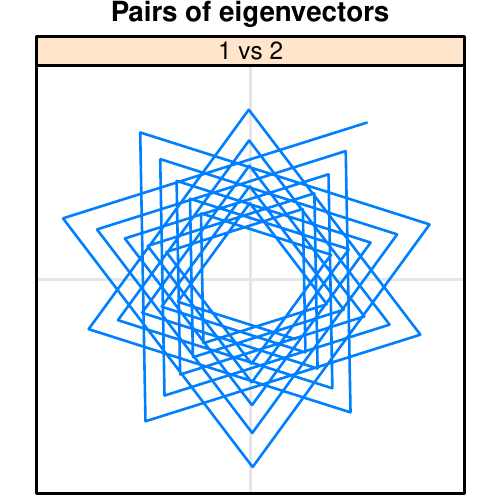}
       \caption{$\omega=3/10$, $\alpha = 0.02$,\\ $C = 1.98$, $\tau =$0.2e-03.}
    \end{subfigure}
    \caption{Two-dimensional diagrams of singular vectors of real e.m. harmonics, $N=99$, $L=50$,     $\beta=1/2$.}
    \label{fig:re_ex_mod_diagram}
\end{figure}

Therefore, in conditions of Proposition~\ref{th:th_tau_1d} for $d=2$ and sufficiently large $L$, the series \eqref{eq:ex_mod_garm_1d} can be identified with a two-dimensional diagram of singular vectors $U_1$ and $U_2$ and the measure $\tau$ will be close to zero. Examples of resulting images with <<spirals>> for series with $N=99$, $L=50$ are presented in Fig.~\ref{fig:re_ex_mod_diagram}.

Under the conditions of Proposition \ref{th:th_tau_1d} for $d=2$,  at sufficiently large $L$, the series \eqref{eq:ex_mod_garm_1d} can be identified with the help of a two-dimensional diagram of singular vectors $U_1$ and $U_2$, which corresponds to the measure $\tau$ close to zero. Fig. \ref{fig:re_ex_mod_diagram} contains several examples of resulting images with <<spirals>> for series with $N=99$, $L=50$.

Note that in practice we do not deal with series of infinite length, so for a series of finite length $N$ with a constant value $\alpha$ the condition $\alpha = C/N$ means not too large value $e^{\alpha N}$, i.e. the limited range of series values.

If the number of e-m harmonics we want to identify is known, the algorithm consists simply in selecting pairs of vectors $U_i$ and $U_{i+1}$ with minimal values of the measure $\tau(U_i, U_{i+1})$, $i=1,\ldots, d-1$.

In the case where the number of e-m harmonics is unknown, pairs of singular vector pairs can be selected by using a threshold, i.e., those vectors $U_i$ and $U_{i+1}$ whose value $\tau(U_i, U_{i+1})$ is less than a given threshold can be assigned to a harmonic, but there is no theoretical justification for choosing the threshold.
Therefore, in this section, we present an empirical justification for the choice of the threshold and conduct numerical studies using the example of a noisy e-m cosine.

Note that the method does not work for the case $\omega = 0.5$. Therefore, we do not consider it in this section.

We will call this method \textit{the method for identifying the oscillatory component by the regularity of angles}.

\subsubsection{Algorithm}
All the explanations given in the previous section explained why the value of the measure $\tau$ will be 0 for the singular vectors of the e-m harmonic. The question arises whether a value $\tau$ close to 0 can be obtained for singular vectors of not the e-m harmonic.

For example, if some vectors $P$ and $Q$ give, although different, small values of the angles $\theta_k$ given in the definition \eqref{eq:tau1} of the measure $\tau$, then the variance $\hat{\Var}(\Theta)$ from the same definition will also be small.
To overcome this problem, we will use a normalized version of the measure $\tau$ for the vectors $P = (p_1,\ldots,p_L)^{\mathrm{T}}$ and $Q=(q_1,\ldots,q_L)^{\mathrm{T}}$:
\begin{gather} \label{eq:tau_norm}
\tilde{\tau}(P, Q) := \frac{\tau(P, Q)}{\min(1, \bar{\theta}^2)} =\frac{\hat{\Var}(\Theta)}{\min(1, \bar{\theta}^2)} =\frac{1}{(L-1) \min(1, \bar{\theta}^2)}\sum_{k=1}^{L-1}{\left(\theta_k  - \bar{\theta}\right)^2},
\end{gather}
where, as before, $\Theta=(\theta_1,\ldots,\theta_L)^{\mathrm{T}}$,
$\bar{\theta} = \sum_{k=1}^{L-1}{\theta_k}/(L-1)$ and $\theta_k$ is the angle between
$\left(p_k, q_{k}\right)^{\mathrm{T}}$ and $\left(p_{k+1}, q_{k+1}\right)^{\mathrm{T}}$.

Dividing by $\min(1, \bar{\theta}^2)$ can only increase the value of the measure, so it fights the problem of small angle values. Small values can be also produced by angles for singular vectors of the e-m harmonic; however, then the variance will be close to zero and normalization will not play a significant role.

For $\tilde{\tau}$ as well as for $\tau$, the statement \ref{th:th_tau_1d} holds, since if $\tau(P, Q) = 0$, then obviously also $\tilde{\tau}(P, Q) = 0$.

As discussed in the previous section, the method has two modifications: for the case where the number of e-m harmonics is known and for the case where the number of e-m harmonics is unknown. These two modifications differ in the stopping criteria at the end of the algorithm \ref{alg:1dtau_2} presented below.

\begin{algorithm}{1D-SSA. The angle-regularity identification method for the oscillatory component}{alg:1dtau_2}
\Input The following data and parameters are input.
\begin{enumerate}
\item \textbf{Data}: number of components $r$; left singular vectors $\{U_j\}_{j=1}^{r}$.
\item \textbf{Parameters}: number of e-m harmonics $m \leq r/2$, or threshold $t_0 \geq 0$, depending on the stopping criterion.
\end{enumerate}
\Output A group of $J$ singular vector indices related to the oscillatory component.
\State Based on $\{U_i\}_{i=1}^{r}$, compute and order in ascending order the values $\tilde{\tau}(U_j, U_{j+1})$, $j=1,\ldots, r-1$ using \eqref{eq:tau_norm}.
While calculating for each $j = 2,\ldots,r-1$ we check: if $\tilde{\tau}(U_j, U_{j+1}) < \tilde{\tau}(U_{j-1}, U_{j})$, then we drop $\tilde{\tau}(U_{j-1}, U_{j})$ from consideration; otherwise we drop $\tilde{\tau}(U_j, U_{j+1})$.
 The resulting values are $\tau_1, \ldots, \tau_{\lfloor r/2 \rfloor}$.
\State Two versions of the stopping criterion are as follows. Try the elements of the set $i=1,\ldots,\lfloor r/2 \rfloor$
\begin{enumerate}
\item either until $i < m$,
\item or until $\tau_i < t_0$.
\end{enumerate}
Denote $i_0$ the moment of stopping.
\State The set $J$ consists of the indices $j$, $j+1$ of the singular vectors $U_j$ and $U_{j+1}$ involved in computing the values $\tau_1, \ldots, \tau_{i_0 - 1}$.
\end{algorithm}

\begin{remark}
Not only the $r$ leading singular vectors can be the input of the algorithm. Any set of consecutive vectors (a set where the index of each next vector is one more than the previous one) from the set $\{U_j\}_{j=1}^{d}$ can be the input. In the algorithm, the variant with the $r$ leading vectors was given for simplicity.
\end{remark}

\begin{remark}
The question about the choice of normalization in the definition of \eqref{eq:tau_norm} needs further elaboration.
%Perhaps with all factors taken into account, the normalization can be chosen better.
\end{remark}

\subsubsection{Choice of the threshold}
\label{sec:treshold_selection}

There are two options for using methods of automatic identification.
\begin{enumerate}
\item Analyze one time series without information in advance.
\item Analyze many time series of similar structures (batch processing).
\end{enumerate}

For the second option, the method of selecting the threshold is straightforward.
For several series from the batch, we compute the values of the measure $\tau$ for consecutive pairs of singular vectors;
look at the images of singular vectors, choose a group of indices $I$ of the components that belong to the oscillatory component; choose a threshold $t_0$ at some point, between the maximum value $\tau$ for $I$ and the minimum value $\tau$ for the remaining indices $\{1,\ldots,d\} \not \in I$. Then we use the obtained value of the threshold $t_0$ for all remaining series. In the \ref{sec:compare} section, the method under consideration will be compared to the frequency method just for this case.

For the first option, it is not possible to invent and theoretically justify a universal method for selecting the threshold. Therefore, the problem of selecting a threshold that leads to extracting only the oscillatory component is replaced by a preprocessing that finds all the oscillatory components, but can also extract something else.

Therefore, let us run simulations of series with a single e-m harmonic and noise; then look at the $95\%$ quantile of the value $\tau(U_1, U_2)$ for different noise levels and the dependence of quantiles on noise levels.

Hereafter, we consider the following $\tS$, $\tN$, and $\tX$ series for model data studies:
\begin{gather} \label{eq:series_S_N}
s_k =  e^{\alpha k} \cos(2\pi k \omega), \\  \label{eq:series_N_N}
n_k =  e^{\alpha k} \sigma \varepsilon_k, \\ \label{eq:series_X_N}
x_k = s_k + n_k,
\end{gather}
where $k=1,\ldots,N$, $\varepsilon_k$ are independent Gaussian random variables distributed as $N(0,1)$. Consider $N = 99$, $L = 50$, $\sigma = 0, 0.2, 0.4, 0.6, 0.8, 1, 1.2, 1.4$. The number of simulations is $1000$.

%Р‘СѓРґРµРј РёСЃРїРѕР»СЊР·РѕРІР°С‚СЊ РґР»СЏ РјРѕРґРµР»РёСЂРѕРІР°РЅРёСЏ С‚Рµ Р¶Рµ СЂСЏРґС‹ $\tS$, $\tN$ Рё $\tX$ СЃ СЌР»РµРјРµРЅС‚Р°РјРё РІРёРґР°
%\eqref{eq:series_S_N}, \eqref{eq:series_N_N}, \eqref{eq:series_X_N}, С‡С‚Рѕ Рё СЂР°РЅРµРµ РІ СЂР°Р·РґРµР»Рµ \ref{sec:tau_study}.
%$N = 99$, $L = 50$, $\sigma = 0, 0.2, 0.4, 0.6, 0.8, 1, 1.2, 1.4$. РњРѕРґРµР»РёСЂРѕРІР°С‚СЊ Р±СѓРґРµРј $1000$ СЂР°Р·.
%$L\omega$ --- РЅРµ С†РµР»РѕРµ, РЅРѕ РєР°Рє РјС‹ СѓР¶Рµ Р·РЅР°РµРј, РґР»СЏ СЂР°СЃСЃРјР°С‚СЂРёРІР°РµРјРѕРіРѕ РјРµС‚РѕРґР° СЌС‚Рѕ РЅРµРІР°Р¶РЅРѕ.

\begin{figure}[!hhh]
	\begin{center}
	\includegraphics[width = 4.5in]{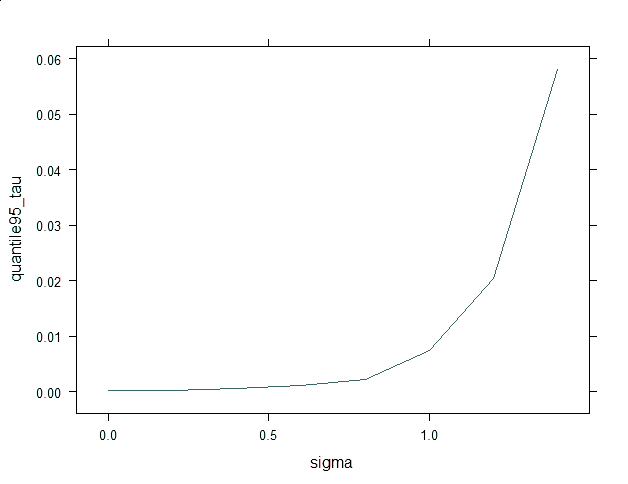}
	\end{center}
	\caption{Dependence of the $95\%$ quantile for $\tau(U_1, U_2)$ on the $\sigma$ value of noise.}
	\label{fig:q95_tau1}
\end{figure}

For each value of $\sigma$, simulate the series $\tN$ $1000$ times and count the $95\%$ quantile of the sample value of the measure $\tau(U_1, U_2)$ from the first two singular vectors based on the obtained sample series in the form \eqref{eq:series_X_N}. Figure ~\ref{fig:q95_tau1} shows a graph of the dependence of the quantile on the value of $\sigma$.

The value $\sigma = 1$ is very large for the length of the series under investigation, i.e., the noise is mixed with the oscillatory component. This can be seen in Fig. ~\ref{fig:q95_tau1}: after $\sigma=1$ the graph went sharply upwards. So it is worth considering the results for $\sigma\le 1$.
The results show that the approximate appropriate value of the threshold is $t_0 = 0.01$ for $\sigma = 1$.
%This value was used in the implementation is as the default value. But experiments with real-life data show that the optimal threshold value can fluctuate greatly.

\subsection{Comparison of methods for harmonic identification}
\label{sec:compare}
Let us compare the method by the regularity of angles with the frequency method from Section~\ref{sec:1d_pgram_method}.
\label{sec:comp_tau1_pgram_many_same_series}
As before, we consider the series $\tS$, $\tN$ and $\tX$ with elements given by \eqref{eq:series_S_N}, \eqref{eq:series_N_N} and \eqref{eq:series_X_N}.
%$N = 99$, $L = 50$, we will simulate $1000$ times.

The aim of automatic identification is the same as that of visual identification.
Let us therefore compare the methods in the following way.
We will simulate the e-m harmonic $\tS$ and noise $\tN$
with $\sigma = 0.2, 0.4, 0.6, 0.8, 1$, $\alpha = 0, 0.01$, $\omega = 1/7$. As before, $N = 99$ and $L = 50$ (therefore, $L\omega$ is not integer).
For fixed values of parameters $\alpha$, $\omega$, $\sigma$ we will model the series $\tX = \tS + \tN$. We will call ``visual identification'' the signal reconstruction by two leading eigentriples.

Let $\tX^{(1)}$ and $\tX^{(2)}$ be two realizations of the series $\tX$.
Let $\tS^{(V,1)}$ and $\tS^{(V,2)}$ be the series reconstructed from the series $\tX^{(1)}$ and $\tX^{(2)}$ by means of <<visual identification>>.

Denote the threshold value for the method with $\tau$ by $t_0$, and for the frequency method by $\rho_0$.
Let $\tS^{(A,1,\varepsilon)}$ and $\tS^{(A,1, \rho_0)}$ be the series recovered from the series $\tX^{(1)}$ using automatic identification algorithms with threshold values $t_0$ and $\rho_0$ respectively.
We also denote the set of values $T = \{0,0.01,0.02,\ldots, 0.99,1\}$, i.e., numbers from $0$ to $1$ in steps of $0.01$.

Then we find the values of thresholds $t_0^{opt}$ and $\rho_0^{opt}$ by solving the following minimization problems:
\begin{gather*}
t_0^{opt} = \argmin_{t_0 \in T}{\left(\frac{1}{N}\sum_{k=0}^{N-1}{\left(s_k^{(V,1)} - s_k^{(A,1,t_0)}\right)}^2\right)}, \\
\rho_0^{opt} = \argmin_{\rho_0 \in T}{\left(\frac{1}{N}\sum_{k=0}^{N-1}{\left(s_k^{(V,1)} - s_k^{(A,1,\rho_0)}\right)}    ^2\right)}.
\end{gather*}

Next, let $\tS^{(A,2,t^{opt})}$ and $\tS^{(A,2, \rho_0^{opt})}$ be the series reconstructed from the series $\tX^{(2)}$ using the automatic identification algorithms with threshold values $t_0^{opt}$ and $\rho_0^{opt}$ respectively.

Let us calculate the identification errors for these series as
\begin{gather*}
E_{\tau} = \frac{1}{N}\sum_{k=0}^{N-1}{\left(s_k^{(V,2)} - s_k^{(A,2,t_0^{opt})}\right)^2}, \
E_{\rho} = \frac{1}{N}\sum_{k=0}^{N-1}{\left(s_k^{(V,2)} - s_k^{(A,2,\rho_0^{opt})}\right)^2}.
\end{gather*}

Thus, we calculated the optimal threshold values for one realization of the $\tX$ series, used these optimal threshold values to reconstruct the series over the second realization, and calculated the identification error for the resulting series.

Let us repeat the above procedure $200$ times and calculate the mean and median of the identification errors $E_{\tau}$ and $E_{\rho}$. The results for $\alpha = 0$, $\omega = 1/7$ are shown in Table~\ref{tab:comp_tau1_pgram}. The results for $\alpha = 0.02$, $\omega = 1/7$ are presented in Table~\ref{tab:comp_tau1_pgram2}.

\begin{table}[hhh!]
\centering
\caption{Comparison of harmonic identification; $\alpha = 0$, $\omega = 1/7$.}
\begin{tabular}{rrrrr}
  \hline
 & mean\_$\tau$ & mean\_$\rho$ & median\_$\tau$ & median\_$\rho$ \\
  \hline
$\sigma$ = 0.2 & 0.0042 & 0.040 & 0 & 0.0028 \\
  $\sigma$ = 0.4 & 0.0156 & 0.064 & 0 & 0.0046 \\
  $\sigma$ = 0.6 & 0.0444 & 0.100 & 0 & 0.0342 \\
  $\sigma$ = 0.8 & 0.1012 & 0.151 & 0 & 0.0656 \\
  $\sigma$ = 1 & 0.1224 & 0.183 & 0 & 0.1179 \\
   \hline
\end{tabular}
\label{tab:comp_tau1_pgram}
\end{table}

\begin{table}[hhh!]
\centering
\caption{Comparison of harmonic identification; $\alpha = 0.02$, $\omega = 1/7$.}
\begin{tabular}{rrrrr}
  \hline
 & mean\_$\tau$ & mean\_$\rho$ & median\_$\tau$ & median\_$\rho$ \\
  \hline
$\sigma$ = 0.2 & 0.0003 & 0.050 & 0 & 0.0010 \\
  $\sigma$ = 0.4 & 0.0010 & 0.080 & 0 & 0.0020 \\
  $\sigma$ = 0.6 & 0.0082 & 0.103 & 0 & 0.0170 \\
  $\sigma$ = 0.8 & 0.0549 & 0.185 & 0 & 0.0655 \\
  $\sigma$ = 1 & 0.1917 & 0.239 & 0 & 0.1554 \\
   \hline
\end{tabular}
\label{tab:comp_tau1_pgram2}
\end{table}

The following conclusions can be drawn from tables \ref{tab:comp_tau1_pgram} and \ref{tab:comp_tau1_pgram2}. First, for all values of $\alpha$ and $\sigma$, the angle regularity identification method with the $\tau$ measure gives smaller errors than the frequency method with the $\rho$ measure.
 Second, both methods give approximately the same results for unmodulated ($\alpha = 0$) and modulated harmonics ($\alpha=0.02$). For the method of identification by the regularity of angles, this result is confirmed by theory, since the measure $\tau$ does not depend on $\alpha$. For the frequency method, this result is unexpected, since the method is based on the periodogram of singular vectors, and for the modulated harmonic there is a ``leakage'' of the periodogram. The result can be explained by the fact that, for the frequency method, the algorithm takes into account possible leakage, and
as the key measure $\rho$ given by the formula \eqref{eq:rho_ij}, the sum of two neighboring values of the periodogram is taken.

\section{Automation of grouping in MSSA}
\label{sec:auto_mssa}
For MSSA, we propose a generalization of the low-frequency method from Section~\ref{sec:1d_freq_method}, the frequency method presented in Section~\ref{sec:1d_pgram_method} and the method by the regularity of angles described in Section~\ref{sec:tau1} for identifying the oscillatory component.
%For 2D-SSA, we generalize only the low-frequency method.

Consider a multivariate time series $\tX= \left(\tX^{(1)}, \ldots,\tX^{(s)}\right)$, $\tX^{(p)}= \left(x^{(p)}(1),\ldots,x^{(p)}(N_p)\right)$, $p=1,\ldots,s$, $x^{(p)}(i) \in \spR$.

The main feature of MSSA, unlike the real or complex case, is that all three kinds of objects: elementary reconstructed series, left singular vectors, and right singular vectors, have different structures.

It follows from the description of the MSSA algorithm that the reconstructed series have the same form as the original multivariate series, i.e. they are also multivariate series (or, the same, a system of time series).

It follows from Proposition~\ref{th:mssa_vec} for MSSA that the left singular vectors are one-dimensional real-valued time series
 of length $L$.
They describe the general structure of the series $\tX^{(1)}, \ldots, \tX^{(s)}$ (the structure of the column space of the trajectory matrix). They have the same form as singular vectors for a single time series.

 The right singular vectors are also one-dimensional real-valued time series; they describe the structure of the row space of the trajectory matrix; each row consists of elements of all series $\tX^{(1)}, \ldots, \tX^{(s)}$, not just one.
We will use the splitting of the right singular vectors $V_i$ into parts $V_i^{(m)}$, $m=1,\ldots,s$, introduced in \eqref{eq:V_mssa}.

Thereby, the algorithms for MSSA automation have different modifications for different variants of the input objects. Different modifications of the algorithms can produce different results.
%, for example, if the algorithm is applied to right singular vectors, as opposed to left singular vectors, each of $s$ rows is taken into account separately in each vector.

\subsubsection{Low-frequency method for trend identification}
\label{sec:freq_method_mssa}

Let us generalize Algorithm~\ref{alg:freq1d} for trend identification algorithm to the case of multivariate series. Algorithm~\ref{alg:freq1d} can be applied in 1D-SSA either to left or right singular vectors or to elementary reconstructed series in the same manner.
As already mentioned, in MSSA the left singular vectors are one-dimensional real-valued time series of length $L$ and have the same form as singular vectors for one series. That is, the left singular vectors have the same form as in the case of 1D-SSA, and Algorithm~\ref{alg:freq1d} of the low-frequency method can be applied to them in the same way. Algorithm~\ref{alg:freqmssa_1} for this case is given below.

%The right singular vectors are one-dimensional real-valued time series of length $K$ and are represented as \eqref{eq:V_mssa}.
The factor vectors $\{V_i\}$ in MSSA consist of parts
related to each time series separately; that is,
\begin{equation}
\label{eq:V_mssa}
    V_i=\left(
        \begin{array}{c}
            V_i^{(1)}\\
            \vdots\\
            V_i^{(s)}
        \end{array}
    \right),
\end{equation}
where the $p$th factor subvector $V_i^{(p)} \in \spR^{K_p}$  belongs to the row trajectory space of the $p$th series.

We can apply the low frequency method to each parts ${V}_i^{(1)}, \ldots, {V}_i^{(s)}$ of the right singular vector $V_i$.
Consider the measure \eqref{eq:T_measure} for each part ${V}_i^{(1)}, \ldots, {V}_i^{(s)}$ of vector $V_i$, and take the maximum. The proposed Algorithm~\ref{alg:freqmssa_2} is formally described below.

The elementary reconstructed series in the case of MSSA are multivariate time series. The idea of applying the low-frequency method to them is the same as for right singular vectors: in the first step, for each of $s$ series, consider the measure by the formula \eqref{eq:T_measure} and take the maximum value of these measures to use in the second step.
This variant of the method is also given by the algorithm \ref{alg:freqmssa_2}, the only difference is that elementary reconstructed series are taken as inputs.

 \begin{algorithm}{MSSA. Low-frequency method for trending: version with left singular vectors}{alg:freqmssa_1}
\Input The following data and parameters are input.
\begin{enumerate}
\item \textbf{Data}: index group $I \in \{1,\ldots,d\}$; frequency $0 \leq \omega \leq 0.5$; left singular vectors $U_i$, $i \in I$.
\item \textbf{Parameters}: threshold $0 \leq T_0 \leq 1$.
\end{enumerate}
\Output A group of indices $J \subset I$ of components related to the trend.
\State For each vector $U_i$, $i \in I$, we calculate the value $T(U_i; \omega)$ using \eqref{eq:T_measure}.
\State $J$ is the group of indices $i \in I$ such that $T(U_i; \omega) \geq T_0$.
\end{algorithm}

\begin{algorithm}{MSSA. Low-frequency method for trending: version with right singular vectors or elementary reconstructed series}{alg:freqmssa_2}
\Input The following data and parameters are input.
\begin{enumerate}
\item \textbf{Data}: index group $I \in \{1,\ldots,d\}$; frequency $0 \leq \omega \leq 0.5$; series $\tY_i$, $i \in I$:
right singular vectors or elementary reconstructed series.
\item \textbf{Parameters}: threshold $0 \leq T_0 \leq 1$.
\end{enumerate}
\Output A group of indices $J \subset I$ of components related to the trend.
\State For each series $\tY_{i} = (\tY_i^{(1)}, \ldots, \tY_i^{(s)})$, $i \in I$, we calculate the values $T(\tY_i^{(1)}; \omega), \ldots,
T(\tY_i^{(s)}; \omega)$ using \eqref{eq:T_measure}. $T_m := \max \{T(\tY_i^{(1)}; \omega); \ldots;
T(\tY_i^{(s)}; \omega)\}$.
\State $J$ is the group of indices $i \in I$ such that $T_m \geq T_0$.
\end{algorithm}

%\newpage
\subsubsection{Frequency method of identification of the oscillatory component}
In the multivariate case, we will also assume that the oscillatory component of each of the series is the sum of the e-m harmonics, i.e. the elements of the multivariate e.m. harmonic series $\tS$ have the form \eqref{eq:em_mssa}.

In MSSA, the left singular vectors are one-dimensional real-valued time series of length $L$ and have the same form as singular vectors for one series% (see Section~\ref{sec:mssa_theory}).
Therefore, the frequency method described in Section~\ref{sec:1d_pgram_method}
can be applied to them in the same way as in the case of 1D-SSA. Below is the algorithm \ref{alg:mssa_pgram_1} of the frequency method for MSSA as applied to left singular vectors.

\newpage
\begin{algorithm}{MSSA. Frequency method for the oscillatory component: version with left singular vectors}{alg:mssa_pgram_1}
\Input The following data and parameters are input.
\begin{enumerate}
\item \textbf{Data}: left singular vectors $\{U_i\}_{i=1}^{d}$.
\item \textbf{Parameters}: parameter $s_0 \in \spZ_{+}$, threshold $\rho_0 \in [0,1]$.
\end{enumerate}
\Output A group of component indices $J^{(\tP)}$ related to the oscillatory component.
\State Based on $\{U_i\}_{i=1}^{d}$, obtain the index group $J_1^{(\tP)}$ using \eqref{eq:I_1_P} with $s_0$ and the index group $J_2^{(\tP)}$ using \eqref{eq:I_2_P} with $s_0$.
\State Based on $\{U_j\}_{j \in J_1^{(\tP)} \cup J_2^{(\tP)}}$, get the index group $J^{(\tP)}$ using \eqref{eq:pgram_I_p} with $\rho_0$.
\end{algorithm}

As mentioned above, the right singular vectors $V_i$
 have the form \eqref{eq:V_mssa}.
Since the vectors to which we apply the frequency method must have norm 1, together with vectors $V_i^{(p)}$ we consider the normalized vectors $
\widehat{V}_i^{(p)} := \frac{V_i^{(p)}}{\|V_i^{(p)} \|}$.

In the first step of the method, the sets of indices $J_{1,p}^{(\tP)}$, $p=1,\ldots,s$, is calculated:
\begin{gather} \label{eq:I_1_P_mssa}
J_{1,p}^{(\tP)} = \{ i: \quad \theta_{i}^{(p)} \theta_{i+1} ^{(p)} >0, \quad L |\theta_i^{(p)} - \theta_{i+1}^{(p)}| \leq s_0, \quad 1 \leq i \leq d -1 \},
\end{gather}
where $\theta_i^{(p)} = \argmax_{0 < k \leq K_p/2} \{\Pi_{\widehat{V}_i^{(p)}}^{K_p}(k/{K_p})\}$, $p=1,\ldots,s$.
Hereby, for each $p = 1,\ldots,s$,
the sets $J_{1,p}^{(\tP)}$ are computed for each pair of sequential vectors $\widehat{V}_i^{(p)}$ and $\widehat{V}_{i+1}^{(p)}$ of length $K_p$.

Similarly, each singular vector is checked against a harmonic with frequency 0.5:
\begin{gather} \label{eq:I_2_P_mssa}
J_{2,p}^{(\tP)} = \{i: \quad {K_p} |\theta_{i}^{(p)} - 0.5 | \leq s_0, \quad 1 \leq i \leq d \}.
\end{gather}

In the second step, we combine the sets of indices: $J_1^{(\tP)}$ is the union of the sets $\{ J_{1,p}^{(\tP)} \}_{p=1}^{s}$, $J_2^{(\tP)}$ is the union of the sets $\{ J_{2,p}^{(\tP)} \}_{p=1}^{s}$.

For the second step, we introduce the definition
\begin{gather*}
\rho_{i,j} :=  \max_p \{ \max_{0 < k \leq {K_p}/2}{\left(\rho_{\{\widehat{V}_i^{(p)}, \widehat{V}_{j}^{(p)}\}}(k/{K_p}) + \rho_{\{\{\widehat{V}_i^{(p)}, \widehat{V}_{j}^{(p)}\}}((k+1)/{K_p})\right)}, p=1,\ldots, s
 \},
\end{gather*}
where $\widehat{V}_i^{(p)}$ and $\widehat{V}_{j}^{(p)}$ are the vectors with indices from the set $J_1^{(\tP)}$ constructed in the first step.
Definition~\ref{def:rho} of the measure $\rho_A$ is given in Section~\ref{sec:1d_pgram_method}.

To identify the components belonging to a harmonic with frequency 0.5, the measure has the form
\begin{gather*}
\rho_{i} :=  \max_p \{ \max_{0 \leq k \leq {K_p}/2}{\left(\rho_{\{\widehat{V}_i^{(p)}\}}(\lfloor {K_p}/2 \rfloor/{K_p}) + \rho_{\{\{\widehat{V}_i^{(p)}\}}((\lfloor {K_p}/2 \rfloor + 1)/{K_p})\right)}, p=1,\ldots, s
 \},
\end{gather*}
where $i \in J_2^{(\tP)}$.

The final result, see~Algorithm~\ref{alg:mssa_pgram_2}, are the indices
\begin{gather} \label{eq:pgram_I_p_mssa_2}
J^{(\tP)} = \{(i,j): \rho_{i,j} \geq\rho_0 \} \cup \{ i: \rho_{i} \geq\rho_0 \}.
\end{gather}

\begin{algorithm}{MSSA. Frequency method for the oscillatory component: version with right singular vectors}{alg:mssa_pgram_2}
\Input The following data and parameters are input.
\begin{enumerate}
\item \textbf{Data}: right singular vectors $\{V_i\}_{i=1}^{d}$.
\item \textbf{Parameters}: parameter $s_0 \in \spZ_{+}$, threshold $\rho_0 \in [0,1]$.
\end{enumerate}
\Output A group of indices $J^{(\tP)}$ of component related to the oscillatory component.
\State
Based on $\{V_i\}_{i=1}^{d}$, obtain the index group $J_{1,p}^{(\tP)}$, $p=1,\ldots,s$, using \eqref{eq:I_1_P_mssa} with $s_0$ and the index group $J_{2,p}^{(\tP)}$, $p=1,\ldots,s$, using \eqref{eq:I_2_P_mssa} with $s_0$;
$J_1^{(\tP)}$ is the union of the sets $\{ J_{1,p}^{(\tP)} \}_{p=1}^{s}$, $J_2^{(\tP)}$ is union of the sets $\{ J_{2,p}^{(\tP)} \}_{p=1}^{s}$.
\State The index group $J^{(\tP)}$ by $\{V_j\}_{j \in J_1^{(\tP)} \cup J_2^{(\tP)}}$ is obtained using \eqref{eq:pgram_I_p_mssa_2} with $\rho_0$.
\end{algorithm}

\subsubsection{Method for identifying the oscillatory component by the regularity of angles}
Again, if we are considering left singular vectors, the algorithm can be applied in the same way as in the 1D-SSA case (Section~\ref{sec:tau1}). The algorithm \ref{alg:tau_mssa} for this case is given below.

\begin{algorithm}{MSSA. The angle regularity identification method for the oscillatory component: the left singular vector variant}{alg:tau_mssa}
\Input The following data and parameters are input.
\begin{enumerate}
\item \textbf{Data}: number of components $r$; left singular vectors $\{U_j\}_{j=1}^{r}$.\
\item \textbf{Parameters}: number of e-m harmonics $m \leq r/2$, or threshold $t_0 \geq 0$, depending on the stopping criterion.
\end{enumerate}
\Output A group of $J$ singular vector indices related to the oscillatory component.
\State Based on $\{U_i\}_{i=1}^{r}$, compute the values $\tilde{\tau}(U_j, U_{j+1})$, $j=1,\ldots, r-1$ using \eqref{eq:tau_norm} and order them in ascending order.
While calculating for each $j = 2,\ldots,r-1$ we check if $\tilde{\tau}(U_j, U_{j+1}) < \tilde{\tau}(U_{j-1}, U_{j})$, then we drop $\tilde{\tau}(U_{j-1}, U_{j})$ from consideration; otherwise, we drop $\tilde{\tau}(U_j, U_{j+1})$.
 The resulting values are $\tau_1, \ldots, \tau_{\lfloor r/2 \rfloor}$.
\State Two versions of the stopping criterion are as follows. Try the elements of the set $i=1,\ldots,\lfloor r/2 \rfloor$
\begin{enumerate}
\item either until $i < m$;
\item or until $\tau_i < t_0$.
\end{enumerate}
Denote $i_0$ the moment of stopping.
\State
The set $J$ consists of the indices  $j$, $j+1$ of singular vectors $U_j$ and $U_{j+1}$ involved in the computation of the values $\tau_1, \ldots, \tau_{i_0 - 1}$.
\end{algorithm}

The method can be applied to the right singular vectors in the same way as it was for the frequency method.
We calculate the values $\tilde{\tau}(V_j^{(p)}, V_{j+1}^{(p)})$, $p=1,\ldots,s$, for a consecutive pair of right singular vectors $V_{j} = (V_j^{(1)} \ldots, V_j^{(s)})$ and $V_{j+1} = (V_{j+1}^{(1)}, \ldots, V_{j+1}^{(s)})$, $j=1,\ldots,r-1$, whose form is defined by the formula \eqref{eq:V_mssa}. Then for each $j$ we take the minimum value $\tilde{\tau}^{(\min)}(V_j, V_{j+1}) =\min_p \{ \tilde{\tau}(V_j^{(p)}, V_{j+1}^{(p)}), p=1,\ldots,s \}$ and to the already obtained values $\tilde{\tau}^{(\min)}(V_j, V_{j+1})$ we apply the standard algorithm as in the one-dimensional case. The resulting algorithm \ref{alg:tau_mssa_fac} is shown below.

\begin{algorithm}{MSSA. Method of identification by the regularity of angles for the oscillatory component: version with right singular vectors}{alg:tau_mssa_fac}
\Input The following data and parameters are input.
\begin{enumerate}
\item \textbf{Data}: number of components $r$; right singular vectors $\{V_j\}_{j=1}^{r}$.\
\item \textbf{Parameters}: number of e-m harmonics $m \leq r/2$; or threshold $t_0 \geq 0$, depending on the stopping criterion.
\end{enumerate}
\Output A group of indices of singular vectors $J$ related to the oscillatory component.
\State
For each pair of vectors $V_{j} = (V_j^{(1)}, \ldots, V_j^{(s)})$ and $V_{j+1} = (V_{j+1}^{(1)}, \ldots, V_{j+1}^{(s)})$
compute the values $\tilde{\tau}(V_j^{(p)}, V_{j+1}^{(p)})$ using \eqref{eq:tau_norm}, $j=1,\ldots, r-1$, $p=1,\ldots,s$
\State Compute $\tilde{\tau}^{(\min)}(V_j, V_{j+1}) =\min_p \{ \tilde{\tau}(V_j^{(p)}, V_{j+1}^{(p)}), p=1,\ldots,s \}$.
\State For each $j = 2,\ldots,r-1$, if $\tilde{\tau}^{(\min)}(V_j, V_{j+1})< \tilde{\tau}^{(\min)}(V_{j-1}, V_{j})$, then we drop the $\tilde{\tau}^{(\min)}(V_{j-1}, V_{j})$ from consideration; otherwise, we drop $\tilde{\tau}^{(\min)}(V_j, V_{j+1})$.
 Arrange these values in ascending order and obtain $\tau_1, \ldots, \tau_{\lfloor r/2 \rfloor}$.
\State Two versions of the stopping criterion are as follows. Try the elements of the set $i=1,\ldots,\lfloor r/2 \rfloor$
\begin{enumerate}
\item either until $i < m$;
\item or until $\tau_i < t_0$.
\end{enumerate}
Denote $i_0$ the moment of stopping.
\State
The set $J$ consists of the indices  $j$, $j+1$ of singular vectors $U_j$ and $U_{j+1}$ involved in the computation of the values $\tau_1, \ldots, \tau_{i_0 - 1}$.
\end{algorithm}

\section{Automation of 2D-SSA}
\label{sec:auto2D}
%In this section, we give generalizations of some methods from Section~\ref{sec:auto1D} for $\tX$ object types different from one-dimensional real-valued time series.

We consider a field of size $N_x \times N_y$: $\tX= \left(x_{i,j} \right)_{i,j=1}^{N_x,N_y}$, $x_{i,j} \in \spR$.

For 2D-SSA, we generalize only the low-frequency method described by Algorithm~\ref{alg:freq1d} for 1D-SSA in Section~\ref{sec:1d_freq_method}.

\paragraph{Low-frequency method for pattern identification}
\label{sec:freq_method_2d}

Define a two-dimensional periodogram for the field $\tY=\left(y_{n,m}\right)_{n,m=1}^{M_x, M_y}$ as
\begin{gather*}
 \Pi_\tY^{M_x M_y} \left(\frac{k}{M_x}, \frac{l}{M_y}\right) = M_x M_y \|G_{kl}\|,
\end{gather*}
where $1 \leq k \leq M_x$, $1 \leq l \leq M_y$, $G_{kl}$ are complex numbers, which are the coefficient of the two-dimensional Fourier expansion of the field $\tY$:
\begin{gather*}
y_{n,m}=\sum_{k=1}^{M_x}\sum_{l=1}^{M_y} G_{kl}\, e^{2\pi\unit \left(nl/M_x + mk/M_y\right)}, \\\
 G_{kl}= \frac{1}{M_x M_y} \sum_{n=q}^{M_x}\sum_{m=2}^{M_y} y_{n,m}\, e^{-2\pi\unit \left(nl/M_x + mk/M_y\right)}.
\end{gather*}

As well as in the 1D-SSA case, we define for the field $\tY$ and $-0.5 \leq \omega_{1}, \omega_{2} \leq 0.5$ measure
\begin{gather}
\label{eq:T_measure_2d}
T(\tY; \omega_{1}; \omega_{2}) = \sum_{k: 0 \leq k/M_x \leq \omega_{1}} \sum_{l: 0 \leq l/M_y \leq \omega_{2}}  I_\tY^{M_x M_y}(k/M_x, l/M_y),
\end{gather}
where $I_\tY^{M_x M_y}(k/M_x, l/M_y) =\Pi_\tY^{M_x M_y} \left(\frac{k}{M_x}, \frac{l}{M_y}\right) / \|\tY\|^2$.

Since $\|\tY\|^2 = \sum_{k=1}^{\lfloor M_x/2 \rfloor}  \sum_{l=1}^{\lfloor M_y/2 \rfloor}  \Pi_{\tY}^{M_x M_y} \left(\frac{k}{M_x}, \frac{l}{M_y}\right)$, then the measure $T(\tY; \omega_{1}; \omega_{2}) $ can be considered as the contribution of frequencies contained in the frequency rectangle $\{[0, \omega_{1} \times [0, \omega_{2}) \}$.

Algorithm~\ref{alg:freq2d} implements the low-frequency method for 2D-SSA.

 \begin{algorithm}{2D-SSA. Low frequency method for trend}{alg:freq2d}
\Input The following data and parameters are input.
\begin{enumerate}
\item \textbf{Data}: index group $I \in \{1,\ldots,d\}$; values $-0.5 \leq \omega_1, \omega_2 \leq 0.5$; fields $\tY_i$, $i \in I$, which can be elementary reconstructed arrays or devectorized left/right singular vectors.
\item \textbf{Parameters}: threshold $0 \leq T_0 \leq 1$.
\end{enumerate}
\Output A group of indices $J \subset I$ of components related to the trend.
\State For each field $\tY_i$, $i \in I$, calculate the value $T(\tY_i; \omega_{1}; \omega_{2})$ using \eqref{eq:T_measure_2d}.
\State The set $J$ is the group of indices $i \in I$ such that $T(\tY_i; \omega_{1}; \omega_{2}) \geq T_0$.
\end{algorithm}

%\section*{Conclusion}

\bibliographystyle{plain}
%\bibliography{../bib/scopus, ../bib/rssa, ../bib/mcssa, ../bib/bio, ../review/reviewssa, ../Deconvolution/rssaw,

\begin{thebibliography}{10}

\bibitem{Alexandrov2006en}
Th. Alexandrov.
\newblock {\em Development of a software package for automatic selection and
  prediction of additive components of time series within the
  ``{Caterpillar}''-{SSA} approach}.
\newblock Phd thesis, St.Petersburg State University, St.Petersburg, 2006.
\newblock In Russian.

\bibitem{Alexandrov2009}
Th. Alexandrov.
\newblock A method of trend extraction using singular spectrum analysis.
\newblock {\em RevStat}, 7(1):1--22, 2009.

\bibitem{Alexandrov.Golyandina2004en}
Th. Alexandrov and N.~Golyandina.
\newblock Automation of extraction for trend and periodic time series
  components within the method ``{Caterpillar}''-{SSA}.
\newblock {\em Exponenta Pro (Mathematics in applications)}, (3--4
  (7--8)):54--61, 2004.
\newblock In Russian.

\bibitem{Alexandrov.Golyandina2005}
Th. Alexandrov and N.~Golyandina.
\newblock Automatic extraction and forecast of time series cyclic components
  within the framework of {SSA}.
\newblock In {\em Proceedings of the 5th St.Petersburg Workshop on Simulation},
  pages 45--50. St. Petersburg State University, 2005.

\bibitem{AlvarezMeza.etal2013}
Andr{\'{e}}s~Marino {\'{A}}lvarez{-}Meza, Carlos~Daniel Acosta{-}Medina, and
  Germ{\'{a}}n Castellanos{-}Dom{\'{\i}}nguez.
\newblock Automatic singular spectrum analysis for time-series decomposition.
\newblock In {\em 21st European Symposium on Artificial Neural Networks,
  {ESANN} 2013, Bruges, Belgium, April 24-26, 2013}, 2013.

\bibitem{Bogalo.etal2021}
Juan Bógalo, Pilar Poncela, and Eva Senra.
\newblock Circulant singular spectrum analysis: A new automated procedure for
  signal extraction.
\newblock {\em Signal Processing}, 179:107824, 2021.

\bibitem{Fisher1929}
R.~A. Fisher.
\newblock Tests of significance in harmonic analysis.
\newblock {\em Proceedings of the Royal Society A}, 125:54--59, 1929.

\bibitem{Golyandina.etal2018}
N.~Golyandina, A.~Korobeynikov, and A.~Zhigljavsky.
\newblock {\em Singular Spectrum Analysis with {R}}.
\newblock Springer-Verlag Berlin Heidelberg, 2018.

\bibitem{Golyandina.etal2003}
N.~Golyandina, V.~Nekrutkin, and D.~Stepanov.
\newblock Variants of the `{C}aterpillar'-{SSA} method for analysis of
  multidimensional time series.
\newblock In {\em Proceedings of II International Conference on System
  Identification and Control Problems (SICPRO 03)}, pages 2139--2168. Moscow:
  V.A.Trapeznikov institute of Control Sciences, 2003.
\newblock (In Russian).

\bibitem{Golyandina.etal2001}
N.~Golyandina, V.~Nekrutkin, and A.~Zhigljavsky.
\newblock {\em Analysis of Time Series Structure: {SSA} and Related
  Techniques}.
\newblock Chapman\&Hall/CRC, 2001.

\bibitem{Golyandina.Zhigljavsky2020book}
N.~Golyandina and A.~Zhigljavsky.
\newblock {\em {S}ingular {S}pectrum {A}nalysis for Time Series}.
\newblock Springer Briefs in Statistics. Springer, 2nd edition, 2020.

\bibitem{Harmouche.etal2018}
J.~Harmouche, D.~Fourer, F.~Auger, P.~Borgnat, and P.~Flandrin.
\newblock The sliding singular spectrum analysis: A data-driven nonstationary
  signal decomposition tool.
\newblock {\em IEEE Transactions on Signal Processing}, 66(1):251--263, Jan
  2018.

\bibitem{Jain.etal2020}
Sahil Jain, Rohan Panda, and Rajesh~Kumar Tripathy.
\newblock Multivariate sliding-mode singular spectrum analysis for the
  decomposition of multisensor time series.
\newblock {\em IEEE Sensors Letters}, 4:1--4, 2020.

\bibitem{Kalantari.Hassani2019}
Mahdi Kalantari and Hossein Hassani.
\newblock Automatic grouping in singular spectrum analysis.
\newblock {\em Forecasting}, 1(1):189--204, 2019.

\bibitem{Vautard.etal1992}
R.~Vautard, P.~Yiou, and M.~Ghil.
\newblock {S}ingular-{S}pectrum {A}nalysis: A toolkit for short, noisy chaotic
  signals.
\newblock {\em Physica~D}, 58:95--126, 1992.

\end{thebibliography}
%../bib/eodus,../bib/eop, ../AutoSSA/biblio, autossa}

\section{Appendix}
\label{sec:append}
\subsection{Decomposition of e-m harmonics in SSA}
\label{sec:1d_ssa_theory}

Since we consider the oscillatory component as a sum of exponentially-modulated harmonics, let us describe the properties of the decomposition of such harmonics.% for all variants of SSA-like algorithms.

The elements of a real-valued exponential-modulated harmonic series $\tS$ have the form
\begin{equation} \label{eq:ex_mod_garm_1d}
 s_n = a\, e^{\alpha n} \cos(2\pi\omega n + \phi),
\end{equation}
where $0 \leq \phi < 2\pi$, $a \not = 0$, $0<\omega \leq 0.5$, $\sin \phi \not = 0$ for $\omega = 0.5$.

As before, $L$ is the window length, $K=N-L+1$, $\Lambda^{(1)}$ and $\Lambda^{(2)}$ are the subspaces of rows and columns of the trajectory matrix of the series $\tS_N$.

$L$-Rank $d$ of the series is the rank of its trajectory matrix and, therefore, the number of left or right singular vectors corresponding to nonzero singular numbers of the trajectory matrix $\mathbf{X}$.

\begin{proposition} \cite[Section 5.1]{Golyandina.etal2001} \label{th:ex_mod_1d}
\begin{enumerate}
\item $L$-Rank $d$ of a series $\tS$ whose elements are \eqref{eq:ex_mod_garm_1d} is $1$ if $\omega = 0.5$; in other cases $d=2$.
\item If $\omega \not = 0.5$, then the subspaces $\Lambda^{(1)}$ and $\Lambda^{(2)}$ have bases \\
\begin{gather*}
\{(1, e\cos(2\pi\omega),\ldots,e^{\alpha (L-1)}\\cos(2\pi (L-1) \omega))^\mathrm{T}, \\
(0, e\sin(2\pi\omega),\ldots,e^{\alpha (L-1)}\sin(2\pi (L-1) \omega))^\mathrm{T}\} \quad \text{and} \\
\{(1, e\cos(2\pi\omega),\ldots,e^{\alpha (K-1)}\cos(2\pi (K-1) \omega))^\mathrm{T}, \\
(0, e\sin(2\pi\omega),\ldots,e^{\alpha (K-1)}\sin(2\pi (K-1) \omega))^\mathrm{T}\}
\end{gather*}
respectively.
\item In the case $\omega = 0.5$, the one-dimensional $\Lambda^{(1)}$ and $\Lambda^{(2)}$ are linear spans of vectors $(1,c,\ldots,c^{L-1})^{\mathrm{T}}$ and $(1,c,\ldots,c^{K-1})^{\mathrm{T}}$ respectively, where $c = -e^{\alpha}$.
\end{enumerate}
\end{proposition}

\begin{proposition} \cite[Proposition 2.3]{Golyandina.etal2003} \label{th:1dssa_num}
Let $\alpha = 0$, $\omega \not = 0.5$ and $L\omega$ and $K\omega$ be integers. Then the eigenvalues of the SVD for the series $\tS$ with elements of the form \eqref{eq:ex_mod_garm_1d} are the same and have the form $\lambda_1=\lambda_2=a^2LK/4$.
\end{proposition}

It follows from the statement \ref{th:ex_mod_1d} that for $\omega \not = 0.5$ the series $\tS$ has exactly two left singular vectors $U_1$ and $U_2$ corresponding to nonzero eigenvalues of the trajectory matrix,
and that the elements of vectors $U_1$ and $U_2$ can be represented in the following form:
\begin{gather} \label{eq:vectors_ex_mod_1d}
	u_k^{(1)} = a_1 e^{\alpha k} \cos(2\pi\omega k + \phi_1), \quad u_k^{(2)} = a_2 e^{\alpha k}\cos(2\pi\omega k + \phi_2),
\end{gather}
where $1 \leq k \leq L$, $0 \leq \phi_1, \phi_2 < 2\pi$, $a_1, a_2 \not = 0$.

\begin{proposition} \label{th:aboutU1U2_ex_mod_1d}
Let $\alpha = \alpha_N = C/N$, where $C$ is-- some constant, and $L=[\beta N]$, where $0<\beta<1$. Then for singular vectors $U_1$ and $U_2$ of the series $\tS$ whose elements have the form \eqref{eq:vectors_ex_mod_1d}, the relations $\lim_{L \rightarrow \infty}{|\phi_1-\phi_2|}=\pi/2 \,\,(\mod \pi)$ and $\lim_{L \rightarrow \infty}{(a_1/a_2)} = 1$ are satisfied.
\end{proposition}

\subsection{Decomposition of e-m harmonics in MSSA}

Consider \textit{multidimensional exponentially modulated (e-m) harmonic series} $\tS = \left(\tS^{(1)},
 \ldots,\tS^{(s)}\right)$, where elements of the $p$-th  series has the form
\begin{equation} \label{eq:em_mssa}
 s^{(p)}(n) = e^{\alpha n} a_p\cos(2\pi\omega n + \phi_p),
\end{equation}
$0 \leq\phi_p < 2\pi$, $a_p \not = 0$, $0<\omega \leq 0.5$, $\sin \phi_p \not = 0$ for $\omega = 0.5$, $n=1,\ldots,N_p$, $p=1,\ldots,s$.

Denote $L$ a window length, $K_p = N_p - L + 1$, $K = \sum_{p=1}^{s} K_p$, $\Lambda^{(1)}$ and $\Lambda^{(2)}$ are the subspaces of rows and columns of the trajectory matrix of the series $\tS$.

\begin{proposition}  \cite[Proposition 2.2]{Golyandina.etal2003}
\label{th:mssa_rank} \label{th:mssa_vec}
\begin{enumerate}
\item
$L$-Rank $d$ of the series $\tS$, whose elements are given by \eqref{eq:em_mssa}, is $1$ if $\omega = 0.5$; in other cases $d=2$.
\item If $\omega \ne 0.5$, then the subspace $\Lambda^{(1)}$ has the basis
\begin{gather*}
\{(1, e\cos(2\pi\omega),\ldots,e^{\alpha (L-1)}\cos(2\pi (L-1) \omega))^\mathrm{T}, \\
(0, e\sin(2\pi\omega),\ldots,e^{\alpha (L-1)}\sin(2\pi (L-1) \omega))^\mathrm{T}\}.
\end{gather*}
\item If $\omega \ne 0.5$, then the subspace $\Lambda^{(2)}$ has the basis
\begin{gather*}
\{(c^{(1)}_1,\ldots,c_{K_1}^{(1)}; \ldots; c^{(s)}_1,\ldots,c_{K_s}^{(s)})^\mathrm{T},
(d^{(1)}_1,\ldots,d_{K_1}^{(1)}; \ldots; d^{(s)}_1,\ldots,d_{K_s}^{(s)})^\mathrm{T} \},
\end{gather*}
where
\begin{gather*}
c^{(p)}_j = a_p \cos(2\pi j \omega + \phi_p) \quad \text{and}\quad d^{(p)}_j = a_p \sin(2\pi j \omega + \phi_p), \\ j=1,\ldots,K_p, \quad p=1,\ldots,s.
\end{gather*}
\item In the case $d=1$, the one-dimensional subspaces $\Lambda^{(1)}$ and $\Lambda^{(2)}$ span the vectors $(1,c,\ldots,c^{L-1})^{\mathrm{T}}$ and $(1,c,\ldots,c^{K-1})^{\mathrm{T}}$ respectively, where $c = -e^{\alpha}$.
\end{enumerate}
\end{proposition}

\begin{proposition} \cite[Proposition 2.3]{Golyandina.etal2003}
\label{th:mssa_num}
Let $\alpha = 0$, $\omega \ne 0.5$ and $L\omega$ and $K\omega$ are integers, $s=2$. Then the eigenvalues of the SVD of the trajectory matrix of the two-dimensional series $\tS$, whose elements are given by \eqref{eq:em_mssa}, are the same and have the form $\lambda_1=\lambda_2=(a^2 + b^2)LK/4$.
\end{proposition}

\begin{example}
\label{ex:mssa_ev}
Let
\begin{equation*}
     s_k^{(1)}=A\cos(2\pi \omega k+\varphi_1),\qquad
     s_k^{(2)}=B\cos(2\pi k \omega k+\varphi_2).
\end{equation*}
If $L\omega$ and $K\omega$ are integer, then $(\tS^{(1)}, \tS^{(2)})$ produces two equal
eigenvalues in MSSA: $\lambda_1=\lambda_2=(A^{2}+B^{2})LK/4$.
Note that the series $\tS^{(1)}$ itself produces two eigenvalues
equal to $A^{2}LK/4$.
\end{example}
\begin{proof}
\begin{enumerate}
\item
    Denote $\psi=\vphi_1/2$. Then ${\bf X}=P_1Q_1^{\rm T}+P_2Q_2^{\rm T}$, where
\bea
\begin{array}{ll}
    p_{1k}= \cos(2\pi\omega (k-1)+\psi),&p_{2k}=-\sin(2\pi\omega (k-1)+\psi),\\
    q_{1m}=A\cos(2\pi\omega (m-1)+\psi),&q_{2m}=A\sin(2\pi\omega (m-1)+\psi),
\end{array}
\eea
    for $1\leq k\leq L$, $1\leq m \leq K$.

    Set ${p}=\|P_1\|/\|P_2\|$,  ${q}=\|Q_1\|/\|Q_2\|$,
    $S=\|P_1\|\;\|P_2\|\;\|Q_1\|\;\|Q_2\|$ and
\bea
    {c}_p=\frac{(P_1,P_2)}{\|P_1\|\;\|P_2\|}, \quad{c}_q=\frac{(Q_1,Q_2)}{\|Q_1\|\;\|Q_2\|}.
\eea
    It follows from \cite[Proposition 5.2]{Golyandina.etal2001} that $\widetilde{\lm}_1=\lm_1/S$
    and $\widetilde{\lm}_2=\lm_2/S$ are the eigenvalues of the matrix
\bea
    {\bf B}=\left(
\begin{array}{ll}
    {p}{q}+{c}_p{c}_q & {q}{c}_p+{p}^{-1}{c}_q\cr
    {p}{c}_q+{q}^{-1}{c}_p & {p}^{-1}{q}^{-1}+{c}_p{c}_q
\end{array}
\right).
\eea
    Therefore, $\widetilde{\lm}_1$ and $\widetilde{\lm}_2$ can be found as the roots of
\bea
    {\widetilde{\lm}}^2-({p}{q}+({p}{q})^{-1}+2{c}_p{c}_q)\widetilde{\lm}
    +1+{c}_p^2{c}_q^2-{c}_p^2-{c}_q^2=0.
\eea
    If $L\omega$ and $K\omega$ are integers, then ${c}_p={c}_q=0$,
    $\|P_1\|^2=\|P_2\|^2=L/2$ and $\|Q_1\|^2=\|Q_2\|^2=A^2K/2$.\\
    Thus, we have the quadratic equation ${\widetilde{\lm}}^2-2\widetilde{\lm}+1=0$
    with one multiple root $\widetilde{\lm}_1=\widetilde{\lm}_2=1$.
    Therefore, $\lm_1=\lm_2=A^2LK/4$.

    In the asymptotic case ($N$ tends to infinity) the quadratic equation is ${\widetilde{\lm}}^2-b(N)\widetilde{\lm}+c(N)=0$, where
    the coefficients $b(N)\rightarrow 2$ and $c(N)\rightarrow 1$ for $N\rightarrow\infty$. Since
    $S$ is equivalent to $A^2LK/4$, $\lm_1\rightarrow A^2LK/4$, $\lm_2\rightarrow A^2LK/4$.
\item
    The trajectory matrix of $(S^{(1)},S^{(2)})$ can be expressed as
    ${\bf X}=P_1Q_1^{\rm T}+P_2Q_2^{\rm T}$, where $P_1$ and $P_2$ are the same as in the one-dimensional case and
\bea
    Q_1=
\left(
\begin{array}{l}
    A\cos(2\pi\omega 0+\vphi_1)\\
    \vdots\\
    A\cos(2\pi\omega (K-1)+\vphi_1)\\
    B\cos(2\pi\omega 0+\vphi_2)\\
    \vdots\\
    B\cos(2\pi\omega (K-1)+\vphi_2)
\end{array}
\right),
    Q_2=
\left(
\begin{array}{l}
    A\sin(2\pi\omega 0+\vphi_1)\\
    \vdots\\
    A\sin(2\pi\omega (K-1)+\vphi_1)\\
    B\sin(2\pi\omega 0+\vphi_2)\\
    \vdots\\
    B\sin(2\pi\omega (K-1)+\vphi_2)
\end{array}
\right).
\eea
    Then $\|Q_1\|^2=\|Q_2\|^2=(A^2+B^2)K/2$, if $L\omega$ and $K\omega$ are integer,
    or tend to this number in the asymptotic case.
    The next steps of the proof are the same as for onedimensional time series.
\end{enumerate}
\end{proof}

\end{document}